# Red STORM – Builder C++ code for a new DGFRS detection system

*Yu.S. Tsyganov*

*FLNR, JINR, 141980 Dubna, Russia*

**Abstract**

***RE**al-time **D**etection and **STOR**age of **M**ulty-chain events* Builder C++ code has been designed to operate together with the new DGFRS detection system in FLNR, JINR. It allows to provide event by event storage and real-time search for ER-α sequences both for 32 strip position sensitive PIPS detector (*CANBERRA NV*) or 48x128 strip DSSSD detector (*Micron Semiconductors*). In the last case it operates in parallel with the ORNL digital system. Namely with this code (PIPS detector branch) it has become possible to detect 13 decay chains of Z=117 nuclei in the reaction $^{249}$Bk+$^{48}$Ca → 117+3,4n.

1. **Introduction**

With the development of an "active correlation" method a new epoch starts in the field of detecting of ultra rare decays of superheavy nuclei (SHN) [1-4]. It was namely the Dubna Gas Filled Recoil Separator (DGFRS), the facility which was applied for discovery of new SHN's in $^{48}$Ca induced nuclear reactions [5,6]. With a new DGFRS detection system was putting into operation in a year 2010 new REDSTORM C++ Builder code was designed and successfully applied in the $^{239}$Bk+$^{48}$Ca → 117+2,3,4n complete fusion nuclear reactions [7]. This code is written for two different scenarios. One of them is to operate with 32-strip position sensitive PIPS detector (CANBERRA NV, Belgium), whereas the second one is developed for 48x128 strip DSSSD detector (Micron Scd.'s, UK) [8].

2. **C++ code brief description**

Both scenarios (A, B) of the REDSTORM code contain code fragment which allows to search for a pointer to a potential ER-α correlation sequence in a real-time mode. It means that nearly just after the detection of ER-α energy-time-position correlated chain code provides the stop of the target irradiation process for a short time in order to detect the forthcoming decays in a background free mode. Below, such a fragment is presented for A execution scenario. In the fig.1 a schematic of real-time algorithm is shown.

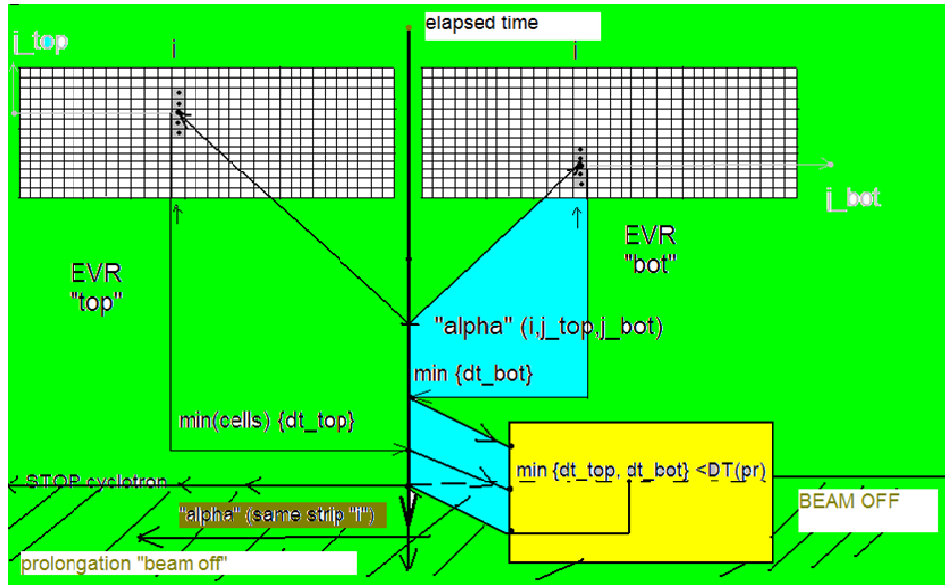

**Fig.1** Schematic of the real-time algorithm. "top" and "bottom" ER matrixes are shown. Beam OFF interval is shown by a dashed area.

*// code fragment for searching for a pointer to ER-alpha correlation*

```
int j=0;

T_PRE_SET=0.00000001;

correlationt=false; str_memo=-1;   pixelt =0; // глоб

pixelt = int (DISCRET*pstT/60.0);

if (pixelt > DIS-4 || pixelt < 4) pixelt=0;

if (EVR== true  && elapsedT > 0.1 ) { RECOT[pixelt][strno]= elapsedT;
    cnt_EVR_top++; }

if (ALFA==true  && situation==false && pixelt > N_PI && pixelt < DIS-N_PI    &&
e_total>EAMIN)
  {
  cnt_top++;

  for (j=0; j < N_PI; j++) dt[j]=0;

  for (j=0; j < N_PI; j++) dt[j]=elapsedT - RECOT[pixelt-N_PI/2 + j][strno];

  dt_min=5.0;

  for (j=0; j<N_PI; j++)

  dt_min =(dt[j]<= dt_min)? dt[j] : dt_min;  /* choice for a minimal time */
```

}....

*//.. code prolongation…*

Note, that a difference between two scenarios (A,B) is achieved by using keys of compilation.

The main difference for B scenario (DSSSD detector) is that only signals from ohmic side of the DSSSD are actual for a real-time process searching for ER-α correlations, whereas junction side signals are used only for determination of number of matrix element of ER-matrix.

Another specific point of the code fragment for B scenario is a necessity to take into account a sharing of signals between any two neighbor strips with probability up to ~20%.

3. **Example of the code application**.

Branch A of REDSTORM code has been successfully applied in the $^{239}$Bk+$^{48}$Ca→117+2,3,4n complete fusion nuclear reactions [7]. Namely with this C++ code the real-time detection of 13 events of Z=117 nucleus has became possible.

In the fig.2 example of Z=117 decaying chains is shown. Signals detected in the Beam OFF phase are shown by the shadow[7].

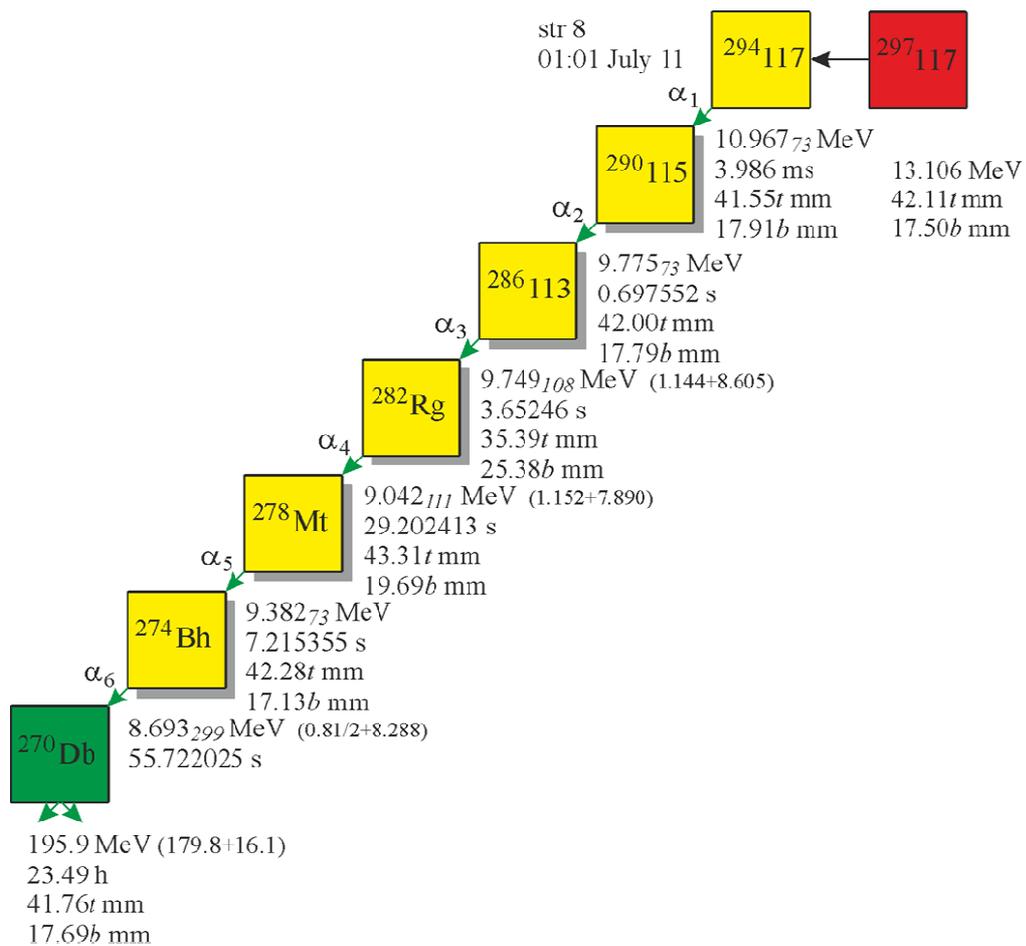

**Fig.2** Detected event of Z=117 nucleus.

## 4. Summary

***RED STORM*** C++ Builder code has been designed for new the DGFRS detection system.
It was applied in the reaction leading to Z=117 element synthesis. The nearest future experiments at the DGFRS are without debts, require of application of "smart" code, which provides a quick search foe ER-α, or even ER-α-α… signal sequences.

## 5. Supplement

When preparing this manuscript good news come from GSI TASCA facility group. Namely, both maim properties of Z=115 and Z=117 nuclei were confirmed in independent experiments [8]. Therefore, way for naming of elements 113,115,117 is, without debts, open now. The same experiment in LBNL (Berkeley, USA) confirmed the DGFRS results too. In a more narrow sense these confirmations indicate to a high quality of the DGFRS detection system.